\documentclass[11pt,a4paper]{article}
\usepackage[utf8]{inputenc}
\usepackage{amsmath}
\usepackage{amsfonts}
\usepackage{amssymb}
\usepackage{amsbsy}

\usepackage{graphicx}
\usepackage[left=3cm,right=3cm,top=2.5cm,bottom=2.5cm]{geometry}
\usepackage{mathrsfs}
\usepackage{tikz}

\usepackage{multirow}
\usepackage{enumitem}

\newcommand{\levy}{L\'{e}vy }
\newcommand{\vx}{\bold{x}}
\newcommand{\vy}{\bold{y}}
\newcommand{\xmin}{x_{\text{min}}}
\newcommand{\xmax}{x_{\text{max}}}
\def\hyph{-\penalty0\hskip0pt\relax}

\title{\textbf{An unsupervised deep learning approach in solving partial integro-differential equations}}
\date{}
\author{Weilong Fu\thanks{Department of IEOR, Columbia University, \texttt{wf2232@columbia.edu}}\ \ \  Ali Hirsa\thanks{Department of IEOR, Columbia University, \texttt{ah2347@columbia.edu}}}

\begin{document}
\maketitle
	
\begin{abstract}
We investigate solving partial integro-differential equations (PIDEs) using unsupervised deep learning in this paper. The PIDE is employed to price options, for the case that the underlying process is a \levy process. In supervised deep learning, pre-calculated labels are used to train neural networks to fit the solution of the PIDE. In unsupervised deep learning, neural networks are employed as the solution, and the derivatives and the integral in the PIDE are calculated based on the neural network. By matching the PIDE and its boundary conditions, the neural network would yield an accurate solution to the PIDE. Once trained, it would be fast for calculating option values as well as option \texttt{Greeks}.
\end{abstract}

\providecommand{\keywords}[1]{\textbf{\textit{Keywords:}} #1}

\section{Introduction}
Financial models based on \levy processes are better at describing the fat tails of asset returns and matching the implied volatility surfaces in option markets than the diffusion models, since \levy processes take jumps into consideration in addition to Gaussian movements. Some examples of the models are the variance gamma model (VG, \cite{madan_variance_1990}), the normal inverse Gaussian model (NIG, \cite{barndorff1997processes}), and the tempered stable process (also known as the CGMY model, \cite{carr_fine_2002}).

The partial integro-differential equation (PIDE) is used to solve the option values under the models based on \levy processes, while the partial differential equation (PDE) is used under the diffusion models. The difference between a PIDE and a PDE is that a PIDE contains an integral term, which is generated by jumps in \levy processes. For this reason, the PIDE is harder to solve and subsequently options under \levy processes are more complex to price. The PIDE can be solved utilizing the finite difference method in an explicit-implicit scheme as described in \cite{hirsa2004pricing}, or the fast Fourier transform (FFT) (see \cite{carr_option_1999} and \cite{lord2008fast} for details). 

Recently, many pricing approaches are proposed based on machine learning (ML) and deep learning (DL). 
\begin{itemize}
	\item In \cite{fu2019fast}, kernel regression is applied to pre-calculated data to price American options. The PIDE is converted into an ordinary integro-differential equation (OIDE), and kernel regression is used to calculate a correction term in the OIDE to reduce pricing errors. 
	\item In supervised deep learning, the neural network is used as a function w.r.t. all the parameters involved in the model. The networks are trained to fit the option price surface or the volatility surface given labels generated by other pricing methods. Recently, this idea has drawn growing attention in the literature (see e.g. \cite{beck2018solving}, \cite{ferguson2018deeply}, \cite{tugce2019}, \cite{liu_neural_2019}, \cite{itkin_deep_2019}, \cite{bayer2019deep}, \cite{horvath2020deep}). The advantage of neural network approaches is that they are fast in computing prices and volatilities once trained and thus they are a good choice for model calibration. However, in supervised learning, it is pretty costly to generate the training labels by other pricing methods, e.g. finite differences, FFT, and/or simulation.
	\item There are unsupervised deep learning approaches as well. The option price surface for a given model is a solution of a PDE or a PIDE and thus the pricing problem is reduced to solving equations. Neural networks have been used to solve PDEs in \cite{lee1990neural}, \cite{lagaris1998artificial}, \cite{lagaris2000neural} and \cite{raissi2018deep}, where the networks are employed as the approximated solutions and the derivatives are calculated either by finite difference or back-propagation \cite{rumelhart1986learning}. The networks are trained to match the PDE and boundary conditions. In this way, the PDE is solved and no labels are needed for training neural networks. Additionally, several modifications are made to deal with high-dimensional problems. In \cite{SIRIGNANO20181339}, the second-order derivatives are estimated by Monte Carlo simulation. In \cite{han2018solving}, the authors made use of forward-backward stochastic differential equations to avoid dealing with second-order derivatives. However, in their method the neural network is used to approximate a term related with the gradient of the solution instead of the solution itself. So far, the literature only considers solving PDEs using the unsupervised deep learning approaches.
\end{itemize}

 The goal of this paper is to extend the unsupervised deep learning approach to the PIDE. In this paper, we propose a pricing method for models based on \levy processes by solving the PIDE with a neural network. The neural network is used as the approximated price surface, and only needs to be trained once, which is the same as supervised deep learning. The main difference from supervised approaches is that this approach is self-contained, which does not need pre-calculated labels. 
 
Also, one benefit of this approach is that the solution given by the neural network yields not only the option price surface but also the Greeks without any extra effort. In comparison, additional labels for Greeks are needed to fit the Greek surface in the supervised approaches. This is easy to understand, since in the supervised approaches, the neural networks are not required to be smooth and there are no constraints of the Greeks during training. While in the unsupervised approaches, the neural networks are required to be smooth for derivative calculation and the derivatives (Greeks) are involved in the PIDE during training.

The paper is organized as follows. In Section \ref{sec:problem}, we describe the model and the PIDE to be solved. The variance gamma model is used as an example for succinctness. In Section \ref{sec:nn}, we introduce the structure of multilayer perceptrons, the activation functions used in this paper, and the process of back-propagation. In Section \ref{sec:cal}, we explain how to calculate the derivatives and the integral in the PIDE. Some further details are provided in Appendix \ref{app:cal}. We also list the boundary conditions and the loss function used for training. In Section \ref{sec:num}, we test the unsupervised deep learning approach for pricing options. We give details of hyper-parameter tuning and assess the effect of boundary conditions and the integral term in the PIDE. At last, we show the results of option prices and Greeks given by the neural network. Section \ref{sec:conclusion} summarizes the paper.

\section{Problem}\label{sec:problem}

\subsection{Model description}
In the paper, we choose the variance gamma (VG) model \cite{madan_variance_1990} as an example for succinctness. The proposed method can be applied to other models based on \levy processes. 

Let $b(t;\theta,\sigma)=\theta t+\sigma W(t)$ be a Brownian motion with drift rate $\theta$ and volatility $\sigma$, where $W(t)$ is a one\hyph dimensional standard Brownian motion. Meanwhile, let $\gamma(t;1,\nu)$ be a gamma process with mean rate $1$ and variance rate $\nu$. Then the three\hyph parameter VG process $X(t;\sigma,\theta,\nu )$ is defined by
\begin{eqnarray*}
X(t;\sigma,\theta,\nu )=b(\gamma(t;1,\nu),\theta,\sigma).
\end{eqnarray*}
The compound process $X(t;\sigma,\theta,\nu )$ can be considered as a time\hyph changed Brownian motion with drift.

The \levy density of the VG process is given by 
\begin{align}
		k(y)=\frac{e^{-\lambda_p y}}{\nu y}1_{y>0}+\frac{e^{-\lambda_n \vert y\vert }}{\nu \vert y\vert}1_{y<0},\label{eq:k}
\end{align}
 where 
 \begin{align*}
 	\lambda_p=\left(\frac{\theta^2}{\sigma^{4}}+\frac{2}{\sigma^2\nu} \right)^{\frac{1}{2}}-\frac{\theta}{\sigma^2}
 \end{align*}
and 
\begin{align*}
\lambda_n=\left(\frac{\theta^2}{\sigma^{4}}+\frac{2}{\sigma^2\nu} \right)^{\frac{1}{2}}+\frac{\theta}{\sigma^2}.
\end{align*} 

The risk neutral process of the stock price under the VG model is given by  
\begin{eqnarray*}
	S(t)=S(0)\exp((r-q)t+X(t)+{\omega} t),
\end{eqnarray*}
where $r$ is the risk\hyph free interest rate, $q$ is the dividend rate of the stock, and $\omega=\frac{1}{v}\ln(1-\sigma^2 \nu/2-\theta \nu)$, where $\omega$ is calculated such that $\mathbb{ E}(S(t))=S(0)\exp((r-q)t)$, i.e., the discounted price process $e^{-(r-q)t}S(t)$ is a martingale. The martingale property of the discounted price is equivalent with the no\hyph arbitrage condition.

\subsection{PIDE}\label{sec:pide}
Suppose $S$ is the stock price, $K$ is the strike price, $t$ is the current time and $T$ is the maturity time, the European put under the VG model is priced by 
\begin{eqnarray*}
	p(S,t)=e^{-r(T-t)}\mathbb{ E}((K-S(T))^{+}\vert S(t)=S). 
\end{eqnarray*}
and the European call is priced by 
\begin{eqnarray*}
	c(S,t)=e^{-r(T-t)}\mathbb{ E}((S(T)-K)^{+}\vert S(t)=S). 
\end{eqnarray*}
Using a martingale approach, one can derive the partial integro-differential equation (PIDE) \cite{hirsa2004pricing}
\begin{eqnarray}
	\int_{-\infty}^{\infty}\left[V(Se^{y},t)-V(S,t)-\frac{\partial V}{\partial S}(S,t)S(e^{y}-1) \right]k(y)dy&& \notag\\
	+\frac{\partial V}{\partial t}(S,t)+(r-q)S\frac{\partial V}{\partial S}(S,t)-rV(S,t)&=&0. \label{eq:pide}
\end{eqnarray}
The prices of European options can be solved by the PIDE with the initial condition
$V(S,T) = (K-S)^{+}$ for put options or $V(S,T) = (S-K)^{+}$ for call options.
Here $k(y)$ is the \levy density as given in \eqref{eq:k}. 

By making change of variables, $x=\ln S$, $\tau=T-t$, and $w(x,\tau)=V(S,t)$, we get 
\begin{eqnarray*}
	\frac{\partial w}{\partial x}(x,\tau)&=&S\frac{\partial V}{\partial S}(S,t),\\
	\frac{\partial w}{\partial \tau}(x,\tau)&=&-\frac{\partial V}{\partial t}(S,t),\\
	w(x+y,\tau)&=&V(S e^{y},t),
\end{eqnarray*}
and the following equation
 \begin{eqnarray}
	\int_{-\infty}^{\infty}\left[w(x+y,\tau)-w(x,\tau)-\frac{\partial w}{\partial x}(x,\tau)(e^{y}-1)  \right]k(y)dy&&\notag \\
	-\frac{\partial w}{\partial \tau}(x,\tau)+(r-q)\frac{\partial w}{\partial x}(x,\tau)-rw(x,\tau)&=&0. \label{eq:pide_middle}
\end{eqnarray}
with the initial condition
$w(x,0) = (K-e^{x})^{+}$ for put options or $w(x,0) = (e^{x}-K)^{+}$ for call options.
Here $x=\log(S)$ is the log-price and $\tau$ is time to maturity. 

Our goal is to solve the PIDE in \eqref{eq:pide_middle} utilizing neural networks directly. We treat the solution $w(x,\tau)$ as a function of $x$ and $\tau$ as well as other parameters, approximate $w(x,\tau)$ with a multi-layer perceptron and train the network to satisfy the PIDE. This is an unsupervised method which means there is no need for labels which are option prices calculated by other methods. For this study we just focus on the European put.

\section{Architecture of the neural network}\label{sec:nn}

Depending on the task and the goal, there are different neural networks that can be utilized. For example, convolutional neural networks (CNNs, \cite{lecun1989backpropagation}) are suitable for image recognition, while recurrent neural networks (RNNs, \cite{hochreiter1997long}) are good at modeling sequential data. For our task, we decide to employ a multi-layer perceptron (MLP), which is a basic kind of neural networks. The numerical results show that the MLP works well and it does not require special architectures to solve the PIDE in an unsupervised approach. 

\begin{figure}[h]
	\begin{tikzpicture}[shorten >=1pt,->,draw=black!50, node distance=2.5 cm]
    \tikzstyle{every pin edge}=[<-,shorten <=1pt]
    \tikzstyle{neuron}=[circle,fill=black!25,minimum size=17pt,inner sep=0pt]
    \tikzstyle{input neuron}=[neuron];
    \tikzstyle{output neuron}=[neuron];
    \tikzstyle{hidden neuron}=[neuron];
    \tikzstyle{annot} = [text width=6em, text centered]

    \foreach \name / \y in {1,...,3}
        \node[input neuron, pin=left:Input \#\y] (I-\name) at (0,-\y) {};

    \foreach \name / \y in {1,...,4}
        \path[yshift=0.5cm]
            node[hidden neuron] (H-\name) at (2.5 cm,-\y cm) {};
    \foreach \name / \y in {1,...,4}
        \path[yshift=0.5cm]
            node[hidden neuron] (J-\name) at (5 cm,-\y cm) {};
            
    \node[output neuron,pin={[pin edge={->}]right:Output}] (O) at (7.5cm,-2) {};

    \foreach \source in {1,...,3}
        \foreach \dest in {1,...,4}
            \path (I-\source) edge (H-\dest);
    
    \foreach \source in {1,...,4}
        \foreach \dest in {1,...,4}
            \path (H-\source) edge (J-\dest);    

    \foreach \source in {1,...,4}
        \path (J-\source) edge (O);

    \node[annot,above of=H-1, node distance=1cm] (hl) {Hidden layer $\vx^{(1)}$};
    \node[annot,left of=hl] {Input layer $\vx^{(0)}$};
    \node[annot,right of=hl] (hl2) {Hidden layer $\vx^{(2)}$};
    \node[annot,right of=hl2] {Output layer $y$};
\end{tikzpicture}
\caption{Illustration of the MLP structure.}
\label{fig:structure}
\end{figure}
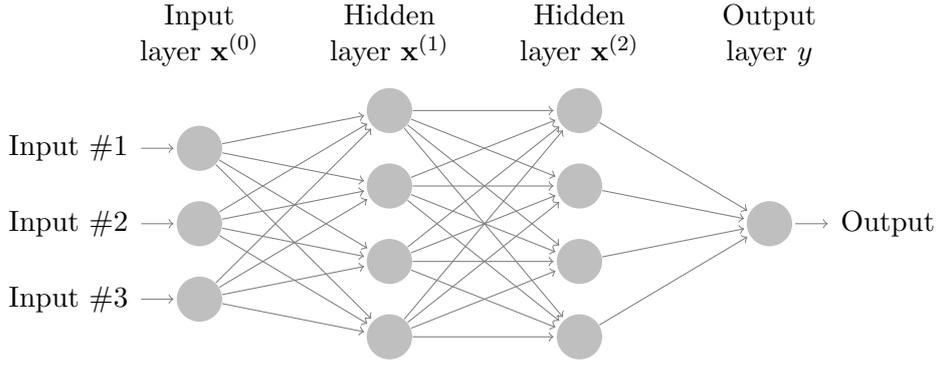

Here we give a simple description of the MLP to keep the paper self-contained. The MLP serves as a multi-dimensional function with an input $\vx\in \mathbb{R}^{n_0}$ and an output $y\in \mathbb{R}$. Suppose the network consists of $L$ hidden layers. Then the MLP can be explained with the equations 
\begin{align*}
	\vx^{(0)}&=\vx, \\
	\vx^{(i)}&=g(W^{(i-1)}\vx^{(i-1)}+b^{(i-1)}), \,\forall 1\leq i\leq L,\\
	y &= W^{(L)}\vx^{(L)}+b^{(L)},
\end{align*}
where the $i$th hidden layer $\vx^{(i)}$ is a vector of length $n_i$ for $1\leq i\leq L$. $n_i$ is the size of the $i$th hidden layer. Though it is possible to let the sizes of the layers be different, we let the sizes be the same for simplicity, i.e., $n_i=n,\forall 1\leq i \leq L$. Thus the dimensions of the parameters are $W^{(0)}\in \mathbb{R}^{n\times n_0}$, $W^{(i)}\in \mathbb{R}^{n\times n}$ for $1\leq i\leq L-1$, $b^{(i)}\in \mathbb{R}^{n}$ for $0\leq i\leq L-1$, $W^{(L)}\in \mathbb{R}^{1\times n}$ and $b^{(L)}\in \mathbb{R}$. A graph of a typical MLP with 2 hidden layers is illustrated in Figure \ref{fig:structure}.

\begin{figure}[h]
\centering
	\includegraphics[width=0.8\textwidth]{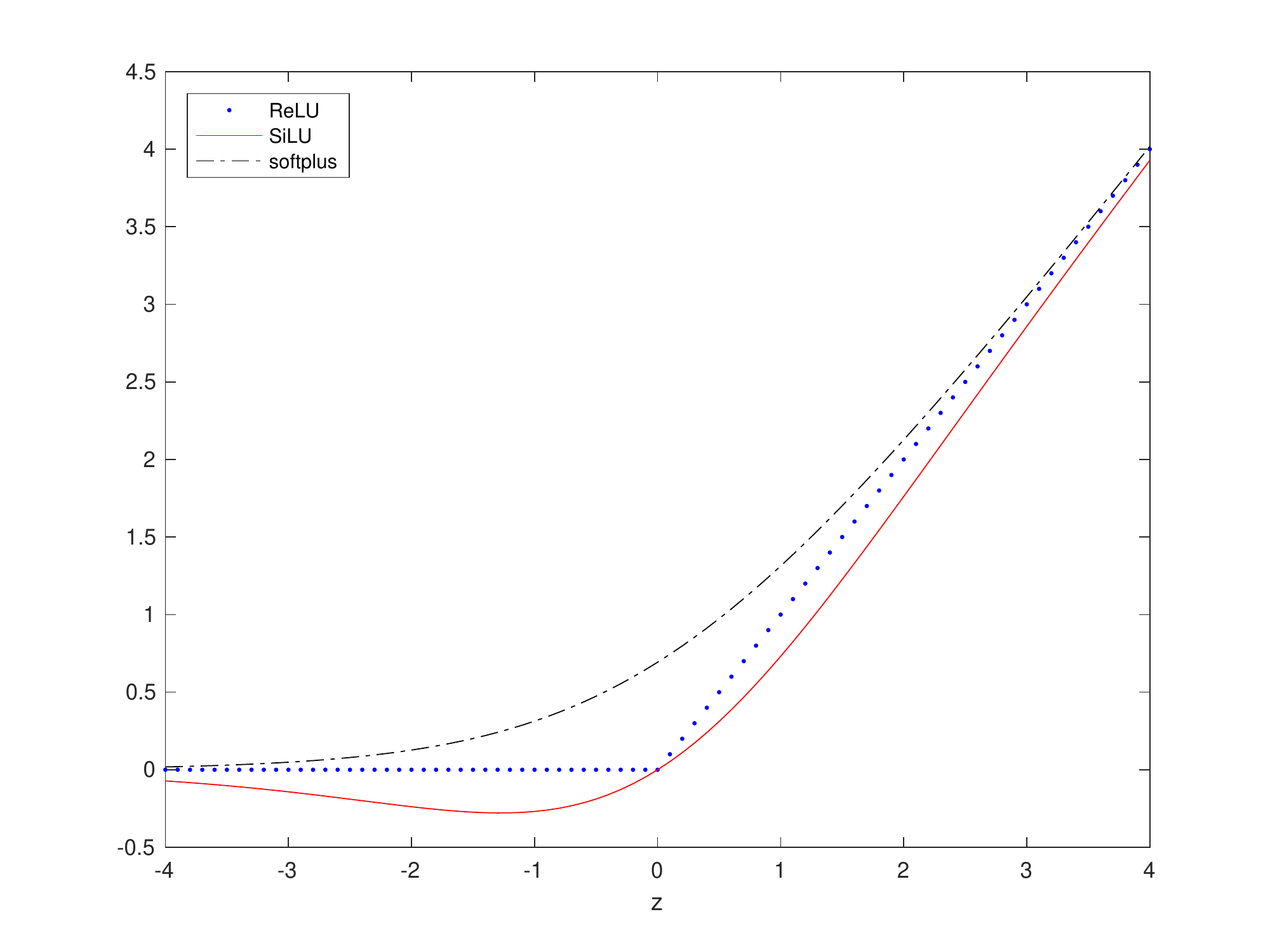}
\caption{Graph of SiLU and softplus compared with ReLU.}
\label{fig:act}
\end{figure}

$g$ is the non-linear activation function which is applied to each coordinate of its input, i.e., $$g(\bold{z})=(g(\bold{z}_1),g(\bold{z}_2),\dots,g(\bold{z}_n)),$$ where $\bold{z}\in \mathbb{R}^{n}$ and $\bold{z}_i,1\leq i\leq n$ are the coordinates of $\bold{z}$. There are some examples of activation functions in Table \ref{tab:act}. The most common used ones are sigmoid, tanh, and ReLU. Sigmoid and tanh are smooth functions. However, their derivatives diminish when the input $z$ tends to infinity. In first order optimization algorithms, diminishing derivatives lead to slow convergence. ReLU does not have a vanishing derivative at infinity, but its derivative is not continuous at 0. Also, an MLP composed of ReLU is always locally linear, which contradicts with the property of the solution $w(x,\tau)$. So we decide to use smooth activations without the problem of vanishing derivatives, e.g., SiLU (also called swish) and softplus. In Figure \ref{fig:act}, we show that SiLU and softplus are two smoothed versions of ReLU. They are different only near the origin.  
\begin{table}[h]
\centering
	\begin{tabular}{|c|c|}\hline
		\textbf{Function}  & \textbf{Definition}\\\hline
		sigmoid & $1/(1+e^{-z})$\\
		tanh & $(e^{z}-e^{-z})/(e^{z}+e^{-z})$\\
		ReLU (\cite{nair2010rectified})& max(0,z)\\
		SiLU (\cite{elfwing2017sigmoidweighted},\cite{ramachandran2017searching} ) & $z/(1+e^{-z})$ \\
		softplus (\cite{dugas2000incorporating}) & $\ln(1+e^{z})$\\\hline
	\end{tabular}
	\caption{Examples of activation functions}
	\label{tab:act}
\end{table}

The output $y$ of the MLP is a composite function of $\vx$ and all the network parameters $W^{(i)},0\leq i\leq L$ and $b^{(i)},0\leq i\leq L$. Given that, the values of $\vx^{(i)},1\leq i\leq L$ and $y$ can be calculated sequentially, which is called forward propagation. The derivatives of $y$ w.r.t. $\vx$, $W^{(i)},0\leq i\leq L$ and $b^{(i)},0\leq i\leq L$ are calculated using the chain rule. The derivative w.r.t. $\vx$ is used to calculate the Greeks in the PIDE and those w.r.t. $W^{(i)},0\leq i\leq L$ and $b^{(i)},0\leq i\leq L$ are used in first-order optimization methods. For neural networks, the chain rule is implemented as back-propagation \cite{rumelhart1986learning}. 

For notation simplicity, we let $$\vy^{(i)} = W^{(i)}\vx^{(i)}+b^{(i)},0\leq i\leq L-1.$$ The back-propagation starts with 
\begin{align*}
	\frac{\partial y}{\partial \vx^{(L)}} &= {W^{(L)}}^{\top}\\
	\frac{\partial y}{\partial W^{(L)}} &= {\vx^{(L)}}^{\top}\\
	\frac{\partial y}{\partial b^{(L)}} &= 1\\
\end{align*}
and is followed by the recursions 
\begin{align*}
	\frac{\partial y}{\partial \vy^{(i)}} &= \frac{\partial y}{\partial \vx^{(i+1)}}\odot g'(\vy^{(i)})\\
	\frac{\partial y}{\partial \vx^{(i)}} &= {W^{(i)}}^{\top} \frac{\partial y}{\partial \vy^{(i)}}\\
	\frac{\partial y}{\partial W^{(i)}} &= \frac{\partial y}{\partial \vy^{(i)}}\,{\vx^{(i)}}^{\top}\\
	\frac{\partial y}{\partial b^{(i)}} &= \frac{\partial y}{\partial \vy^{(i)}}\\
\end{align*}
for $i=L-1,L-2,\dots,1,0$. Here $z^{\top}$ means the transpose of $z$ and $\odot$ means the element-wise product. These steps give the first-order derivatives of $y$ w.r.t. $\vx$, $W^{(i)},0\leq i\leq L$ and $b^{(i)},0\leq i\leq L$. In Section \ref{sec:cal}, we use the first-order and second-order derivatives of the neural network in the loss function \eqref{eq:loss} given in Section \ref{sec:loss}, so the back-propagation in our method is more complex than the above description and is completed by deep learning packages.

\section{Calculation}\label{sec:cal}
\subsection{Derivatives and integral}
In this section, we use an MLP as an approximation of the value function $w(x,\tau)$. The input of the neural network is $\vx=(x,\tau, \sigma,\nu,\theta,r,q)$ and the output is used as the value of $w(x,\tau)$. Let $\xi=\{\sigma,\nu,\theta,r,q\}$. We need to keep in mind that both $w(x,\tau)$ and $k(y)$ are dependent on $\xi$. We omit them in notations for simplicity. Since we always use a smooth activation function to build the neural network, e.g., SiLU or softplus, the neural network is smooth and the derivatives $\frac{\partial w}{\partial \tau}(x,\tau)$, $\frac{\partial w}{\partial x}(x,\tau)$ and $\frac{\partial^2 w}{\partial x^2}(x,\tau)$ (which is used shortly after) are calculated by back-propagation. For a sample $\vx$, we calculate the integral $$\int_{-\infty}^{\infty}\left[w(x+y,\tau)-w(x,\tau)-\frac{\partial w}{\partial x}(x,\tau)(e^{y}-1)  \right]k(y)dy$$ in two parts. The inner part $$\int_{\vert y \vert \leq \epsilon}\left[w(x+y,\tau)-w(x,\tau)-\frac{\partial w}{\partial x}(x,\tau)(e^{y}-1)  \right]k(y)dy$$ is approximated by  
$$\left[\frac{\partial^2 w}{\partial x^2}(x,\tau)-\frac{\partial w}{\partial x}(x,\tau)\right]\int_{\vert y \vert\leq \epsilon}\frac{y^2}{2} k(y)dy$$ the same way as described in Chapter 5 in \cite{hirsa2016computational}.
For the outer part, we write it as \begin{align*}
	&\int_{\vert y \vert> \epsilon}\left[w(x+y,\tau)-w(x,\tau)-\frac{\partial w}{\partial x}(x,\tau)(e^{y}-1)  \right]k(y)dy\\&=\int_{\vert y \vert> \epsilon}\left[w(x+y,\tau)-w(x,\tau)\right]k(y)dy-\frac{\partial w}{\partial x}(x,\tau)\int_{\vert y \vert> \epsilon}(e^{y}-1)  k(y)dy.
\end{align*} The first portion of it is calculated using the trapezoidal rule as explained in \cite{trape}. 
 Further details of the inner and outer part are included in Appendix \ref{app:cal}. With the help of neural network approximation, we can calculate each part on the left hand side of the PIDE \eqref{eq:pide_middle} given a sample $\vx$.

\subsection{Boundary conditions}
For European puts, we have the initial condition given $\vx$: 
\begin{align}
	w(x,0) = (K-e^{x})^{+}\label{eq:bd1}
\end{align} 
For the boundary conditions, we let the value of $x$ in all samples be limited between two fixed boundaries $\xmin$ and $\xmax$. For Dirichlet boundary conditions, we have
\begin{align}
	w(\xmin, \tau) & = K e^{-r\tau}-e^{\xmin-q\tau}\label{eq:bd2}\\
	w(\xmax,\tau) & = 0 \label{eq:bd3}
\end{align}
\subsection{Loss function}\label{sec:loss}
Given a sample $\vx$, the loss function is defined to be the sum of the squared residuals of Equations \eqref{eq:pide_middle}, \eqref{eq:bd1}, \eqref{eq:bd2} and \eqref{eq:bd3}, i.e., 
\begin{align}
	L(\{W^{(i)},b^{(i)}\}_{i=0}^{L};\vx) =& \left(	\int_{-\infty}^{\infty}\left[w(x+y,\tau)-w(x,\tau)-\frac{\partial w}{\partial x}(x,\tau)(e^{y}-1)  \right]k(y)dy\right. \notag \\
	&\left. -\frac{\partial w}{\partial \tau}(x,\tau)+(r-q)\frac{\partial w}{\partial x}(x,\tau)-rw(x,\tau)\right)^2  \notag \\
	&+ \left(w(x,0) - (K-e^{x})^{+}\right)^2 \notag \\
	&+ \left(w(\xmin, \tau) - (K e^{-r\tau}-e^{\xmin-q\tau})\right)^2
	+ \left(w(\xmax,\tau)\right)^2 \label{eq:loss}
\end{align} 
The function $w(x,\tau)$ depends on the network parameter $\{W^{(i)},b^{(i)}\}_{i=0}^{L}$ implicitly since it is approximated by a neural network. For multiple samples, the total loss function is defined as the average of the loss functions on each sample.

\section{Numerical experiments}\label{sec:num}
In the numerical parts, we first introduce the distribution of the samples. Next, we tune the hyper-parameters to select good designs of neural networks. Then we analyze the influence of boundary conditions and the numerical integral on the pricing errors. Finally, we fully train the neural network and show the prices and Greeks given by the neural network.

\subsection{Input dimension}

For VG, the input dimension of the neural network is seven, since $\vx=\{x,\tau,r,q,\theta,\sigma,\nu\}$. In other models based on \levy processes, there would be more or fewer parameters and the input dimension of the neural network would vary accordingly. The input dimension does not influence the performance much since neural networks are good at approximating high-dimensional functions.

\subsection{Settings of parameters and samples}
We consider the option price $w(\cdot)$ in the following region:
\begin{align*}
0< &\tau\leq3,\\ 
1\%\leq &\sigma\leq 50\%,\\
0.1\leq &\nu \leq 0.6,\\ 
-0.5\leq &\theta\leq -0.1,\\ 
0\leq r&,q\leq 0.1
\end{align*} 
The strike $K$ is fixed to 200. 

For the training samples, the log-price $x$ follows the uniform distribution between $\ln(K/40)$ and $\ln(2K)$. For the test samples, the log-price $x$ follows the uniform distribution between $\ln(K/2)$ and $\ln(2K)$. The log-price $x$ of the test samples are restricted between $\ln(K/2)$ and $\ln(2K)$ since we do not want to focus too much on the deep in-the-money options. The other parameters are uniformly distributed within their ranges. The samples are given by the Sobol sequence \cite{sobol1967distribution}, which is a quasi random sequence. 

For the boundary conditions, we assume $x_{min}=\ln(1)$ and $x_{max}=\ln(10000)$. We use $\epsilon=0.01$ when separating the integral into the inner part and the outer part.

\subsection{Hyper-parameter tuning}\label{sec:hyper}
We consider networks consisting of $1\leq L\leq 6$ layers, with $n\in\{100,200,300,400,500,600\}$ neurons in each layer. The activation function is chosen from SiLU and softplus. Initialization changes among He-normal, He-uniform \cite{he2015delving}, LeCun-normal, LeCun-uniform \cite{lecun2012efficient}, Glorot-normal and Glorot-uniform \cite{glorot2010understanding}. The initial distributions are uniform distributions with different ranges or truncated normal distributions with different variances. For example, the He-normal initialization employs a truncated normal distribution with the variance $2/N_{\text{in}}$ where $N_{\text{in}}$ is the input size of the layer. The optimizer could be Adam \cite{kingma2014adam} or RMSprop \cite{tieleman2012lecture}. We also consider regularization of batch-normalization \cite{ioffe2015batch} and dropout \cite{srivastava2014dropout}. There are many choices of the hyper-parameters, and the number of combinations grows exponentially with the number of hyper-parameters. It is not practical to test all combinations. Instead, we analyze the important hyper-parameters step by step.

Here we let the training size be 50,000 and the test size be 2,000. The batch size is 200 and the training epochs is 30. They will be the same from Section \ref{sec:hyper} to \ref{sec:int}.
\begin{enumerate}[label=\arabic*)]
\item \textbf{Initilization}

Given $L=3$ layers and $N=200$ neurons in each layer, we test different initializations. The activation is SiLU and the optimizer is Adam. No regularization is used. 
As shown in Table \ref{tab:init}, He-normal initialization gives the best result, and we choose He-normal going forward in our study.

\item \textbf{Activation}

We compare SiLU and softplus, with the optimizer Adam and no regularization.
As shown in Table \ref{tab:acti}, SiLU performs uniformly better than softplus. We continue with SiLU in the following parts.

\item \textbf{Optimizer}

We compare Adam and RMSprop without any regularization.
As shown in Table \ref{tab:opti}, Adam performs uniformly better than RMSprop. We continue with Adam in the following parts.

\item \textbf{Batch normalization}

We now test batch-normalization.
As shown in Table \ref{tab:batch-norm}, there is no obvious improvement with batch-normalization. Considering batch normalization is costly, we prefer not to use it.

\item \textbf{Number of layers and size of each layer}

We now test the number of layers $L$ and the neuron size $N$ in each layer.
The best combinations are $(L,N)=(4,600)$, $(L,N)=(5,400)$ and $(L,N)=(5,500)$ as shown in Table \ref{tab:size}. We would choose $4\leq L\leq 5$ and $400\leq N\leq 600$. 

\item \textbf{Dropout}

When we choose $L=4$ and $N=400$, the best choice for the dropout rate is 0.3. However, if we choose $L=4$ and $N=200$, the best choice for the dropout rate is 0.2. So the optimal dropout rate choice depends on the size of the network and cannot be fixed. 

Also, a dropout rate slightly larger than 0 does not always reduce the error. We can easily see this phenomenon in Table \ref{tab:drop}. The best performance is reached at a very large dropout rate. There are cases that dropout does not help at all, as can been seen in Tables \ref{tab:bs2} and \ref{tab:big2}. Typically, for certain neural networks, we try different dropout rates and pick the best one. 
\end{enumerate}

\begin{table}[p]
\centering
	\begin{tabular}{|c|c|c|}
	\hline
	& RMSE & MAE\\
	\hline
	Glorot-normal & 2.434 & 13.266\\
 	Glorot-uniform & 2.125 & 10.107\\
 	\textbf{He-normal} & \textbf{1.954} & \textbf{10.140} \\
 	He-uniform & 2.090 & 10.106\\
 	LeCun-normal & 1.956 & 10.433\\
 	LeCun-uniform & 2.129 & 11.486\\
	\hline 
	\end{tabular}
\caption{Comparison of initialization. (RMSE is the root mean squared error and MAE is the maximum absolute error. Same in the following tables.) }
\label{tab:init}
\end{table}

\begin{table}[p]
\centering
	\begin{tabular}{|c|cc|}
	\hline
	& \multicolumn{2}{c|}{RMSE}  \\
	& softplus & SiLU \\
	\hline
$L=3,N=100$ & 2.808 & 2.468  \\
$L=3,N=200$ & 2.309 & 1.954  \\ 
$L=3,N=300$ & 1.861 & 1.614 \\
$L=3,N=400$ & 1.845 & 1.733  \\ 
	\hline 
	\end{tabular}
\caption{Comparison between SiLU and softplus. }\label{tab:acti}
\end{table}

\begin{table}[p]
\centering
	\begin{tabular}{|c|cc|cc|}
	\hline
	& \multicolumn{2}{c}{Adam} & \multicolumn{2}{|c|}{RMSprop}  \\
	& RMSE & MAE & RMSE & MAE \\
	\hline
 $L=3,N=100$ & 2.468 & 11.586 & 4.333 & 14.905 \\
 $L=3,N=200$ & 1.954 & 10.140 & 3.033 & 13.991 \\
 $L=3,N=300$ & 1.614 & 8.569 & 3.075 & 9.670 \\
 $L=3,N=400$ & 1.733 & 10.668 & 2.170 & 10.345 \\
	\hline 
	\end{tabular}
\caption{Comparison between Adam and RMSprop. }\label{tab:opti}
\end{table}

\begin{table}[p]
\centering
	\begin{tabular}{|c|cc|cc|}
	\hline
	& \multicolumn{2}{c}{False} & \multicolumn{2}{|c|}{True}  \\
	& RMSE & MAE & RMSE & MAE \\
	\hline
 $L=3,N=100$ & 2.468 & 11.586 & 2.135 & 10.366 \\
 $L=3,N=200$ & 1.954 & 10.140 & 2.294 & 12.299 \\
 $L=3,N=300$ & 1.614 & 8.569 & 1.888 & 9.882 \\
 $L=3,N=400$ & 1.733 & 10.668 & 1.604 & 9.183 \\
	\hline 
	\end{tabular}
\caption{Performance with (True) and without (False) batch-normalization.}\label{tab:batch-norm}
\end{table}

\begin{table}[p]
\centering
	\begin{tabular}{|cc|ccccccc|}
	\hline
	RMSE & & \multicolumn{7}{c|}{N}  \\
 	& & 100 & 200 & 300 & 400 & 500 & 600 & 700\\
	\hline
 \multirow{4}{*}{L} & 1 & 35.942 & 18.907 & 17.124 & 15.574 & 14.980 & & \\
& 2 & 2.744 & 2.400 & 2.438 & 2.458 & 2.579 & & \\
& 3 & 2.468 & 1.954 & 1.614 & 1.733 & 1.744 & 1.269 & 1.502\\
& 4 & 2.143 & 1.860 & 1.657 & 1.401 & 1.221 & 0.981 & 1.229\\
& 5 & 1.769 & 1.285 & 1.794 & 1.119 & 1.033 & 1.729 & 1.201\\
& 6 & 1.706 & 1.261 & 1.396 & 1.708 & 1.362 & 1.664 & 1.167\\
	\hline 
	\end{tabular}
\caption{Comparison of the number of layers (L) and the layer size (N). }
\label{tab:size}
\end{table}

\begin{table}[p]
\centering
	\begin{tabular}{|c|cc|cc|}
	\hline
	 &\multicolumn{2}{c|}{$L=4,N=400$} &   \multicolumn{2}{c|}{$L=4,N=200$}\\
	\hline
	dropout rate	& RMSE & MAE & RMSE & MAE \\
	\hline
 0 & 1.954 & 10.140 & 1.860 & 6.606\\
 0.1 & 2.092 & 7.114 & 1.708 & 10.718\\
 0.2 & 1.201 & 8.232 & 1.522 & 10.250\\
 0.3 & 0.955 & 4.764 & 1.612 & 10.488\\
 0.4 & 1.211 & 6.760 & 1.635 & 8.638\\
	\hline 
	\end{tabular}
\caption{Effect of dropout.}
\end{table}

\begin{table}[p]
\centering
	\begin{tabular}{|c|cc|cc|}
	\hline
	 &\multicolumn{2}{c|}{$L=4,N=600$} &   \multicolumn{2}{c|}{$L=5,N=500$}\\
	\hline
	dropout rate	& RMSE & MAE & RMSE & MAE \\
	\hline
0.0 & 0.981 & 6.031 & 0.864 & 3.848 \\
0.1 & 0.752 & 3.585 & 1.265 & 8.555 \\
0.2 & 1.512 & 8.228 & 1.227 & 5.744 \\
0.3 & 2.507 & 10.394 & 1.335 & 8.298 \\
0.4 & 1.009 & 6.937 & 1.510 & 8.554 \\
0.5 & 1.894 & 6.381 & 0.742 & 3.886 \\
0.6 & 0.831 & 6.277 & 1.248 & 6.913 \\
0.7 & 1.387 & 8.294 & 1.193 & 6.990 \\
	\hline 
	\end{tabular}
\caption{Effect of dropout. }
\label{tab:drop}
\end{table} 

\subsection{Influence of boundary conditions}
We illustrate in Table \ref{tab:neumann} the result of replacing Dirichlet boundary conditions
\begin{align*}
	w(x_{min}, \tau) & = K e^{-r\tau}-e^{x_{min}-q\tau}\\
	w(x_{max},\tau) & = 0 
\end{align*}
with Neumann boundary conditions 
\begin{align*}
	\frac{\partial^2 w}{\partial x^2} w(x_{min}, \tau) - \frac{\partial w}{\partial x} w(x_{min}, \tau) & = 0\\
	\frac{\partial^2 w}{\partial x^2} w(x_{max}, \tau) - \frac{\partial w}{\partial x} w(x_{max}, \tau) & = 0 
\end{align*}
For small $N$, assuming Neumann conditions, the results are generally better than when considering the Dirichlet conditions. However, for large $N$, the results of the Neumann conditions are worse. Also, as we see from Table \ref{tab:neumann}, the results of the Neumann conditions are less stable and not consistent. There are cases that the results are better and on the other hand there are cases that they are worse. We keep using Dirichlet boundary conditions hereafter.

\begin{table}[p]
\centering
	\begin{tabular}{|c|cc|cc|}
	\hline
	 RMSE &\multicolumn{2}{c|}{Neumann} &   \multicolumn{2}{c|}{Dirichlet}\\
	\hline
	$N$	& $L=4$ & $L=5$ & $L=4$ & $L=5$ \\
	\hline
100 & 1.941 & 1.422 & 2.143 & 1.769 \\
200 & 1.254 & 0.842 & 1.860 & 1.285 \\
300 & 1.512 & 1.109 & 1.657 & 1.794 \\
400 & 2.823 & 0.633 & 1.401 & 1.119 \\
500 & 1.675 & 1.754 & 1.221 & 1.033 \\
	\hline 
	\end{tabular}
\caption{Comparison between Dirichlet and Neumann boundary conditions.}
\label{tab:neumann}
\end{table}

\begin{table}[p]
\centering
	\begin{tabular}{|c|cc|}
	\hline
	$L=4,N=400$, dropout=0.3 &RMSE & MAE \\
	\hline
	original grid & 1.081 & 6.776 \\
	\hline
finer grid & 1.260 & 7.662 \\
	\hline 
	\end{tabular}
\caption{Tests on the error of the trapezoidal rule. }
\label{tab:grid}
\end{table}

\begin{table}[p]
\centering
	\begin{tabular}{|c|cc|}
	\hline
	$L=4,N=400$, dropout=0.3 &RMSE & MAE \\
	\hline
	original method  & 1.081 & 6.776 \\
	\hline
fixed integral & 1.249 &5.611 \\
	\hline 
	\end{tabular}
\caption{Tests of the stability of the numerical integral.}
\label{tab:fixed}
\end{table}

\subsection{Effect of the integral}\label{sec:int}
In this part we analyze the influence of the integral on the pricing errors in three aspects.
\begin{enumerate}[label=\arabic*)]
	\item \textbf{Errors of the numerical integral}
	
	 First we test whether the error of the numerical integral leads to pricing errors. In Appendix \ref{app:cal} we explain how to calculate the integral $$\int_{ y > \epsilon}\left[w(x+y,\tau)-w(x,\tau)  \right]k(y)dy$$ via the trapezoid rule. If we use a finer grid in the trapezoid rule by reducing the gap between the grid points by half, we should expect the error of the numerical integral to be reduced. However, the pricing error gets slightly larger. The performance of the finer grid and the original grid is shown in Table \ref{tab:grid}. So approximation due to usage of the trapezoid rule is not the reason for the pricing error. 
\item \textbf{Stability of updating parameters in the integral}

Next we would like to test whether our methodology is stable as we update the network parameters $\{W^{(i)},b^{(i)}\}_{i=0}^{L}$ in the numerical integral. For a given sample $\vx$, the PIDE \eqref{eq:pide_middle}, as well as the loss function \eqref{eq:loss}, involves not only the local value $w(x,\tau)$, but also the global values $w(x+y,\tau), y \in \mathbb{R}$. In the trapezoidal rule, the global values correspond to $w(x+y_j,\tau),1\leq j\leq J$, where $y_j,1\leq j\leq J$ are the grid points. Hence the loss function of a sample point $\vx$ is dependent on the network parameters not only by the value and derivatives at $(x,\tau)$, but also through the values at many other points $(x+y_j,\tau),1\leq j\leq J$. It is essential to make sure the method is stable by updating the network parameters in both $w(x,\tau)$ and $w(x+y_j,\tau),1\leq j\leq J$ at the same time.

If we compute $w(x+y_j,\tau),1\leq j\leq J$ first and then fix their values while updating the network parameters in each timestep, it means we fix the numerical integral $$\int_{ y > \epsilon}\left[w(x+y,\tau)-w(x,\tau)  \right]k(y)dy$$ in the loss function. In this case, we do not have to take the derivatives of the integral with regard to the network parameters. This is consistent with the explicit-implicit finite difference scheme proposed in \cite{hirsa2004pricing}. This approach is faster, but the error is still larger than the original method. This is easy to understand since we use integral approximation from the last timestep in training to approximate the integrals of the current timestep. Since the fixed integral does not reduce errors, we are not concerned about the stability of updating the network parameters in the numerical integral.

\item\textbf{Comparison with the Black-Merton-Scholes (BMS) model}

To analyze the impact of the integral on the pricing error as a whole, we compare the results of the VG model with those of the BMS model. We consider solving the BMS equation
\begin{eqnarray*}
	-\frac{\partial w}{\partial \tau}(x,\tau)+\frac{\sigma^2}{2}\frac{\partial^2 w}{\partial x^2}(x,\tau)+\left(r-q-\frac{\sigma^2}{2} \right)\frac{\partial w}{\partial x}(x,\tau)-rw(x,\tau)&=&0.
\end{eqnarray*}
using the same neural network structure.
The performance is provided in Tables \ref{tab:bs1} \& \ref{tab:bs2}.

Note that there is no integral term in the BMS equation. If the integral term is the main reason for the errors in the VG model, the errors in the BMS model should be much smaller. But in fact, with neural networks of the same size, the errors in the BMS model are just slightly smaller than those in the VG model. 
Also note that the BMS model is the special case of the VG model when $\nu=0$ and $\theta=0$. The dimension of the sample space in the BMS model is 5, while the dimension in the VG model is 7. The BMS model should be less affected by the curse of dimensionality and we should expect that the errors of neural networks are slightly smaller under the BMS model given the same sample size. 
From these results we can conclude that the integral part is not the main source of the error. 

\end{enumerate}

From the three aspects of the numerical tests on the integral, we conclude that the numerical integral is not the main reason of the pricing errors and it is stable to update the network parameters of the numerical integral in each timestep.

\begin{table}[p]
\centering
	\begin{tabular}{|cc|cccc|}
	\hline
	RMSE & & \multicolumn{4}{c|}{N}  \\
 	& & 300 & 400 & 500 & 600 \\
	\hline
 \multirow{2}{*}{L} & 4 & 1.101 & 0.774 & 1.412 & 0.981\\
            & 5 & 0.955 & 1.100 & 0.684 & 0.882\\
	\hline 
	\end{tabular}
\caption{Results of the BMS model with different network sizes. }
\label{tab:bs1}
\end{table}

\begin{table}[p]
\centering
	\begin{tabular}{|c|cc|}
	\hline
	 &\multicolumn{2}{c|}{$L=5,N=500$} \\
	\hline
	dropout rate	& RMSE & MAE \\
	\hline
0.0 & 0.684 & 3.659 \\
0.1 & 3.721 & 10.742 \\
0.2 & 1.244 & 6.841 \\
0.3 & 1.942 & 6.482 \\
0.4 & 0.960 & 4.942 \\
0.5 & 0.973 & 4.669 \\ 
	\hline 
	\end{tabular}
\caption{Results of the BMS model with different dropout rates. }
\label{tab:bs2}
\end{table}

\begin{table}[p]
\centering
	\begin{tabular}{|c|cc|}
	\hline
	size = 1000000 &RMSE & MAE \\
	\hline 
	$L=5,N=500$, dropout=0 & 0.113 & 0.976 \\
	\hline
	$L=4,N=400$, dropout=0.3 & 0.128 & 1.161 \\
	\hline
$L=4,N=600$, dropout=0 & 0.173 & 1.451 \\
\hline
	\end{tabular}
\caption{Results of fully trained networks. }
\label{tab:big1}
\end{table}

\begin{table}[p]
\centering
	\begin{tabular}{|c|cc|}
	\hline
	size = 1000000 &RMSE & MAE \\
	\hline 
	$L=5,N=500$, dropout=0 & 0.113 & 0.976 \\
	\hline
	$L=5,N=500$, dropout=0.3 & 0.179 & 1.248 \\
	\hline
$L=5,N=500$, dropout=0.5 & 0.226 & 1.417 \\
\hline
	\end{tabular}
\caption{Effect of dropout on fully trained networks. }
\label{tab:big2}
\end{table}

\begin{table}[p]
\centering
	\begin{tabular}{|c|cc|}
	\hline
	$L=5,N=500$, dropout=0 &RMSE & MAE \\
	\hline 
	size = 50000 & 0.111 & 1.136 \\
	
	\hline 
	size = 200000 & 0.125 & 0.970 \\
	\hline
	size = 500000 & 0.142 & 1.163 \\
	\hline
size = 1000000 & 0.113 & 0.976 \\
\hline
	\end{tabular}
\caption{Comparison of different sample sizes. }
\label{tab:big3}
\end{table}

%
%

\subsection{Fully trained networks}\label{sec:full}
The previous tests were to find the suitable neural networks for pricing, and test the performance of the boundary conditions and the numerical integral. We then train the networks on large samples for enough epochs to get the full performance. Tables \ref{tab:big1}-\ref{tab:big3} contain the results of large samples from 50,000 to 1,000,000. The test size is 10,000 in this section. Since the batch size is fixed to 200, there are more timesteps in one epoch if the sample size is larger and more timesteps usually mean better results in optimization algorithms. To be unbiased in our comparison, we need to keep the total timesteps the same even if the sample sizes are different. This is unbiased because the time costs will be the same for different sample sizes. For the size of 50,000, 200,000, 500,000 and 1,000,000, the number of epochs are 600, 150, 60, and 30 respectively. Thus the total timesteps of optimization is $$\frac{1,000,000}{200} \times 30 = 150,000.$$

The neural networks of the best performance are listed in Table \ref{tab:big1}. The RMSE is about 0.1, which is quite small. The MAE is about 1, which is 10 times as large as the RMSE, which means the neural network is quite close to the true price surface, but with a few large meanderings.

In Table \ref{tab:big2}, we show that dropout would not always work. For the sizes such as $L=5,N=500$ or $L=4,N=600$, dropout deteriorates performance. 

In Table \ref{tab:big3}, we fix the structure of the neural network and just change the sample size. When we increase the sample size from 50,000 to 1,000,000, the performance is about the same. 

\subsection{Summary of numerical experiments}
\begin{itemize}
	\item In the numerical parts, we find it is better to use \texttt{He-normal} initialization, SiLU activation, Adam optimizer and no batch-normalization.
	\item  A network of $4\leq L\leq 5$ and $400\leq N\leq 600$ gives better results and the best one in our tests is $L=5$ and $N=500$. 
	\item Dropout would not always reduce errors and the optimal dropout rate should be selected specifically. It seems that no dropout would be a safe choice.
	\item The Dirichlet boundary conditions work well in the method. 
	\item The trapezoid rule works well and the integral term is not the main reason of pricing errors.
	\item The number of total timesteps plays a more important role than the sample size in fully training neural networks. 
\end{itemize}

  
\subsection{Evaluation of prices and Greeks}
\begin{figure}[p]
\centering
	\includegraphics[width=1\textwidth]{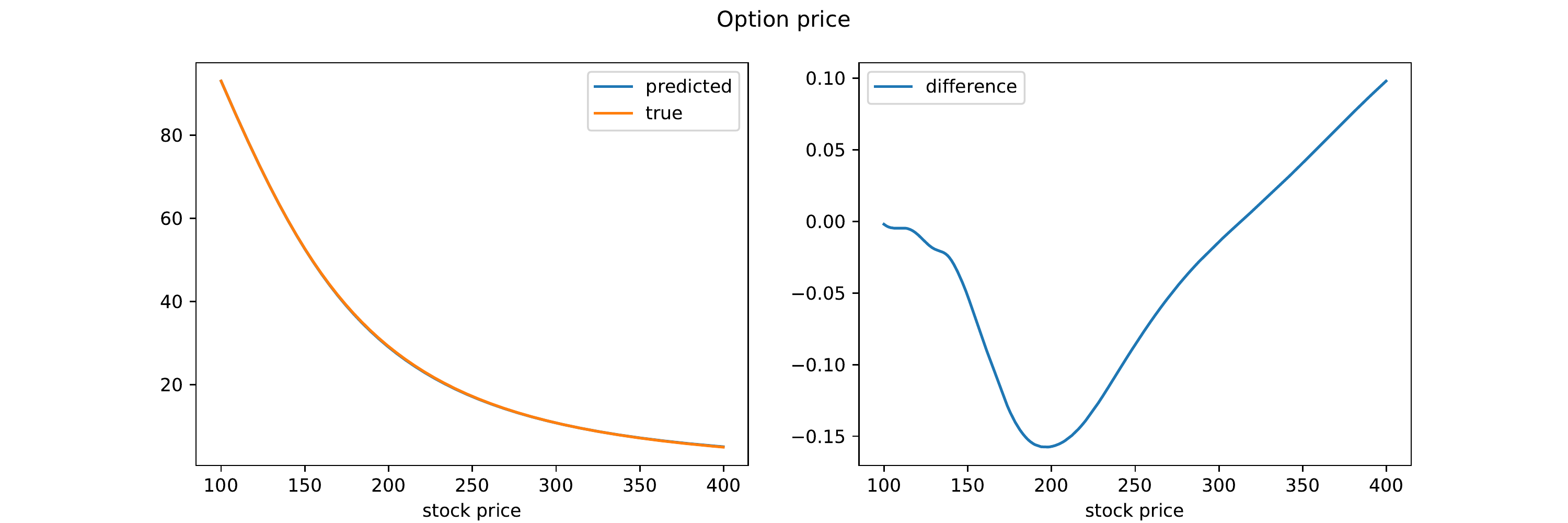}
	\includegraphics[width=1\textwidth]{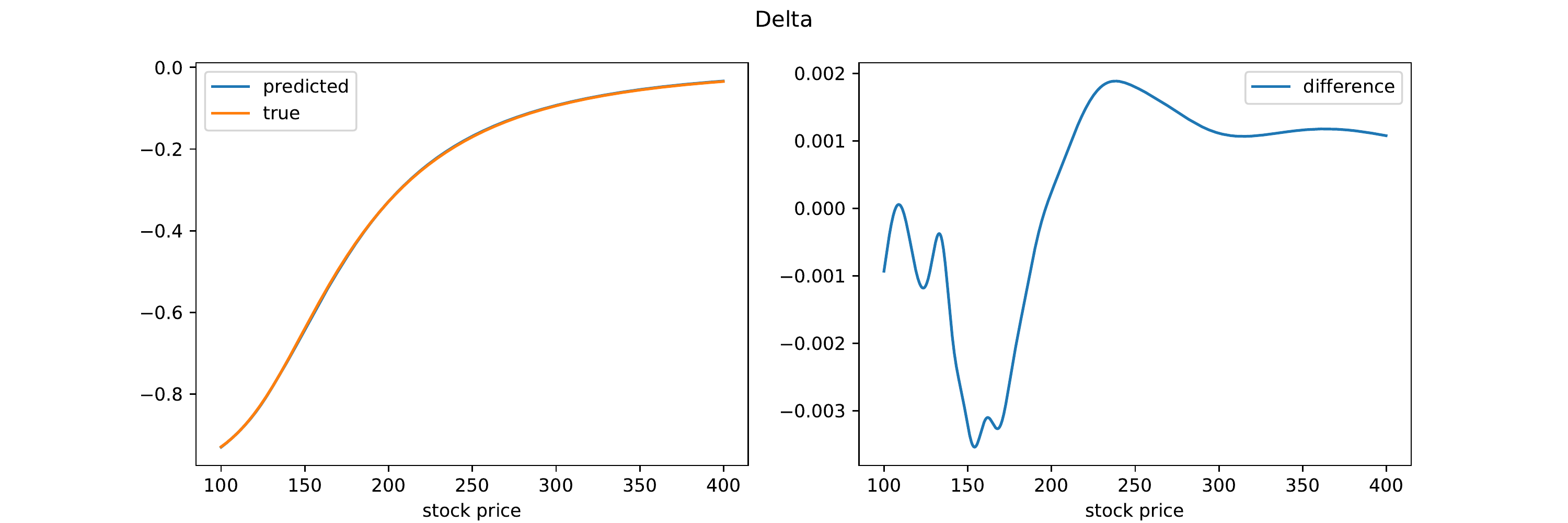}
	\includegraphics[width=1\textwidth]{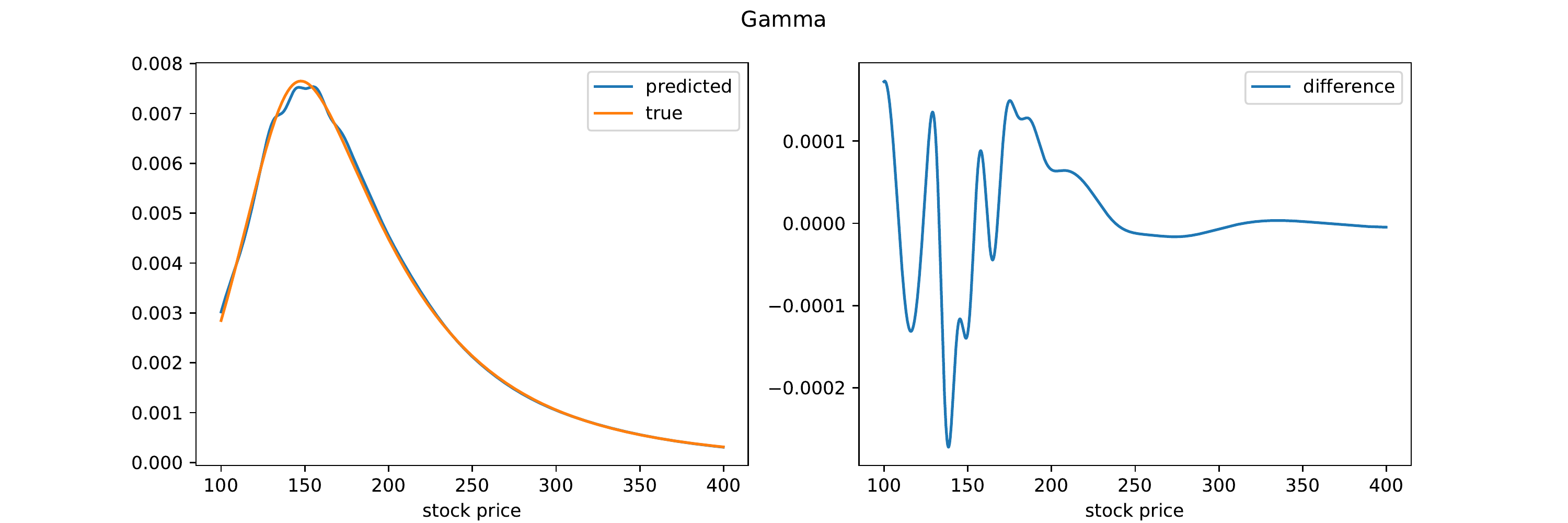}
	\includegraphics[width=1\textwidth]{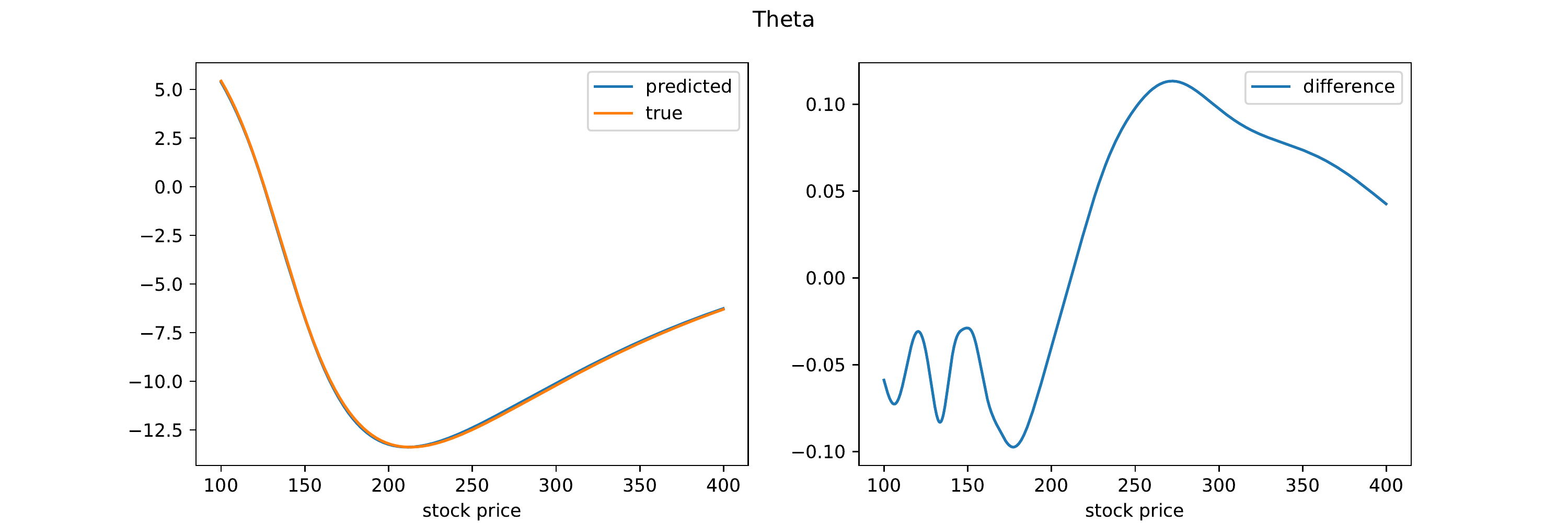}
\caption{Price, delta, gamma and theta for the following parameter set $\{\tau=1$, $r=0.05$, $q=0.02$, $\theta = -0.4$, $\sigma = 0.4$ and $\nu=0.4\}$.}
\label{fig:t1}
\end{figure}

\begin{figure}[p]
\centering
	\includegraphics[width=1\textwidth]{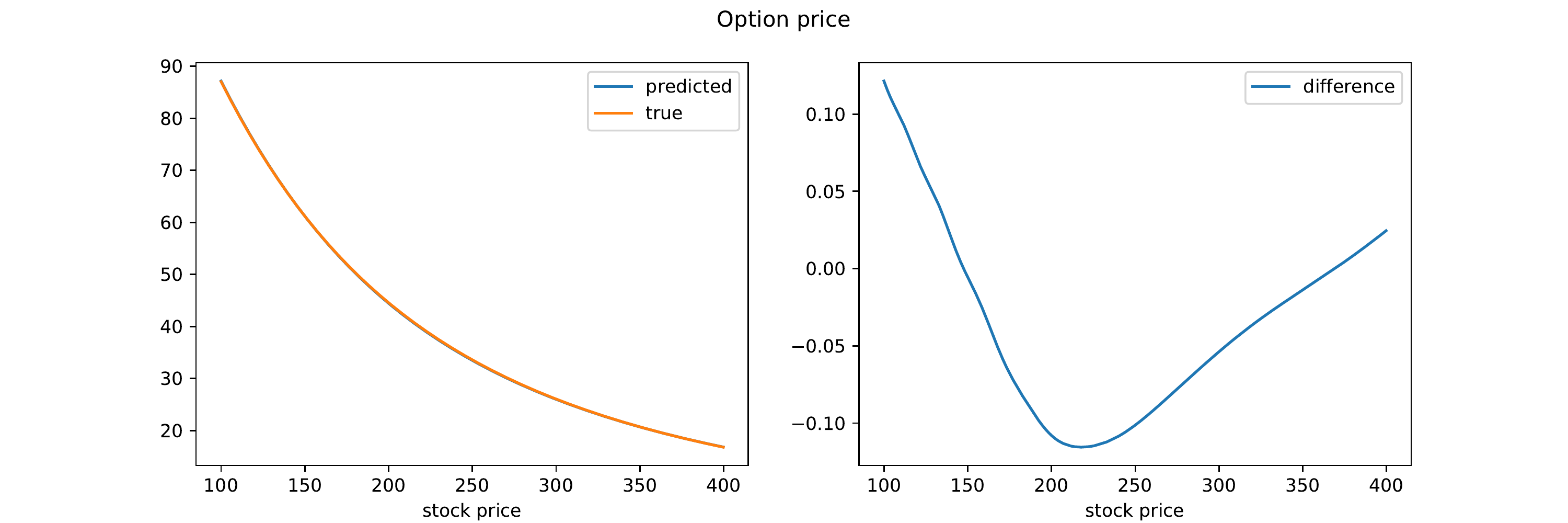}
	\includegraphics[width=1\textwidth]{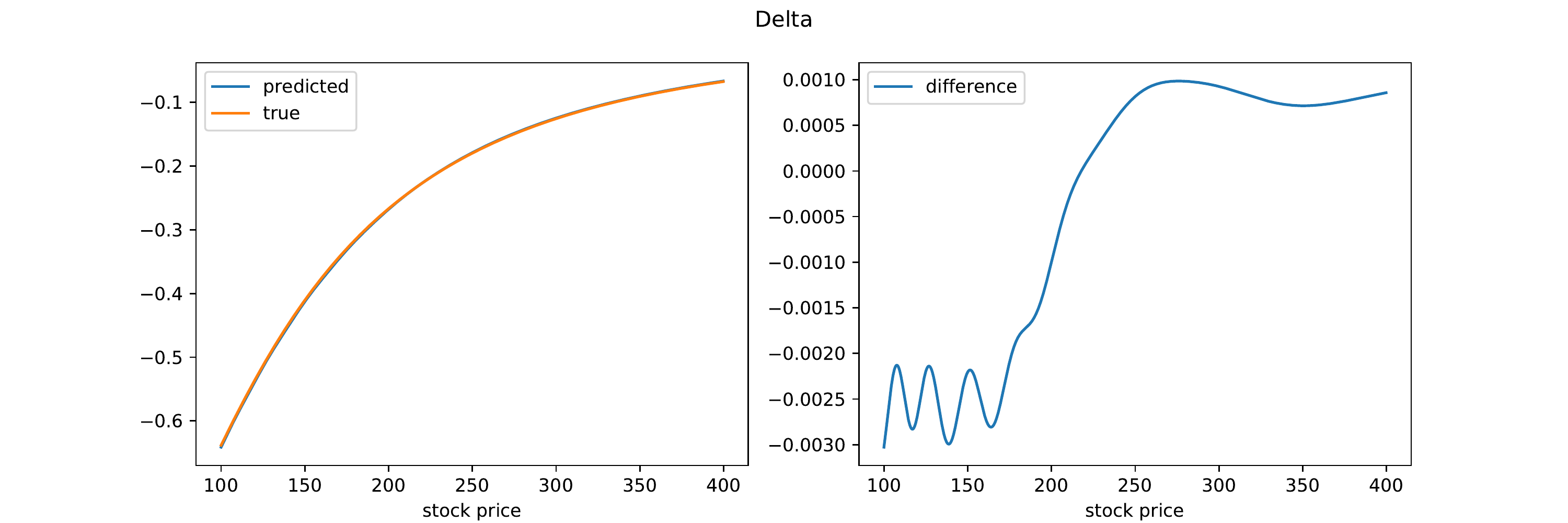}
	\includegraphics[width=1\textwidth]{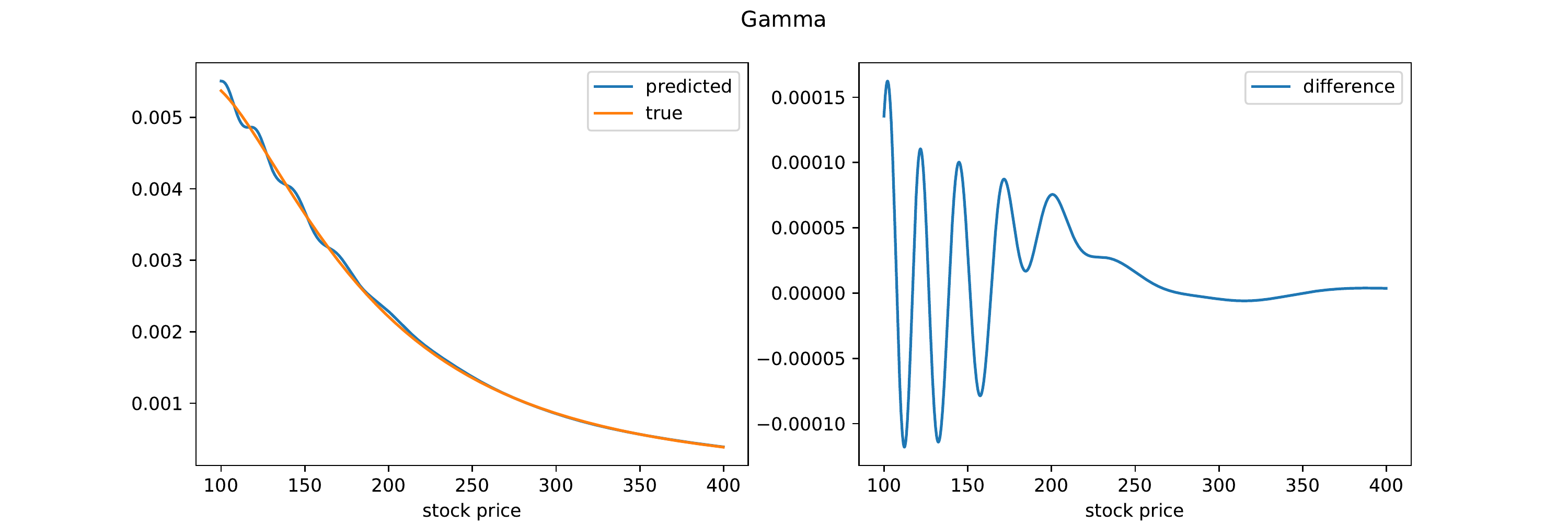}
	\includegraphics[width=1\textwidth]{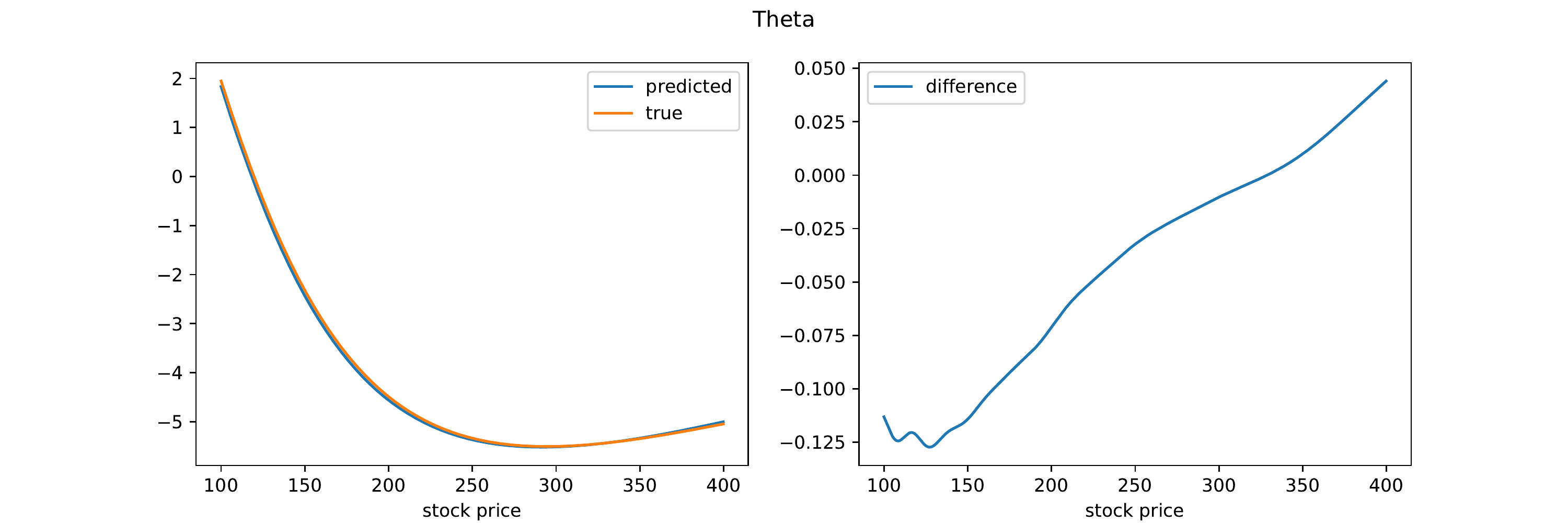}
\caption{Price, delta, gamma and theta for the following parameter set $\{\tau=3$, $r=0.05$, $q=0.02$, $\theta = -0.4$, $\sigma = 0.4$ and $\nu=0.4\}$.}
\label{fig:t3}
\end{figure}

In Figures \ref{fig:t1} and \ref{fig:t3}, we show the curves of the price, delta $\Delta$, gamma $\Gamma$, and theta $\Theta$ and compare the true values and the fitted values from the neural network. The true values are computed through the fast Fourier transform (FFT) \cite{carr_option_1999}.
 The fitted values are calculated from the neural network by back-propagation. Suppose $V(S,t)$ is the option price of a certain strike $K$. Recall the variable changes in Section \ref{sec:pide}. Then delta is $$\Delta = \frac{\partial V}{\partial S}=S^{-1}\frac{\partial w}{\partial x},$$ gamma is $$\Gamma = \frac{\partial^2 V}{\partial S^2}=S^{-2}\left(\frac{\partial^2 w}{\partial x^2}-\frac{\partial w}{\partial x}\right)$$ and theta is $$\Theta = \frac{\partial V}{\partial t} = -\frac{\partial w}{\partial \tau}. $$ The model for evaluation is the network of $L=5,N=500$ and dropout=0 after 30 epochs of training on a sample size of 1,000,000, as given in Section \ref{sec:full} and Table \ref{tab:big1}.

In Figure \ref{fig:t1}, we choose $K=200$, $\tau = 1$ (1 year), $r=0.05$, $q=0.02$, $\theta = -0.4$, $\sigma = 0.4$ and $\nu=0.4$. In Figure \ref{fig:t3}, all the parameters are the same except $\tau=3$ (3 years). Even though we only want to fit an approximate solution of the price, we also get the option Greeks from the neural network. Here we only show the option Greeks w.r.t. price and time. In fact, we can get the Greeks w.r.t. $r$, $q$ and all the model parameters form the neural network by back-propagation.

\section{Conclusion}\label{sec:conclusion}
In this paper we have proposed a pricing approach using unsupervised deep learning. Specifically, we use an MLP to approximate the solution to the PIDE of the VG model. The method can be applied to other models based on \levy processes, if we replace the \levy density $k(y)$ and change the input of the neural network. In the numerical parts, we show that the numerical integral which is particular in PIDE does not increase the pricing errors.

The first benefit of this approach is that we only need to train the neural network once for a given model. The second benefit is that we do not need labels for training. The third is that by this approach we do not only obtain the option price itself, but we also get the option Greeks.

We only study the European options in the paper. For future study, it is attractive to extend the same approach to price American options under \levy processes.
  
\newpage  
\bibliographystyle{abbrv}
\bibliography{references.bib}

\newpage
\appendix
\section*{Appendix}
\section{Calculation of the integral in PIDE}\label{app:cal}
This part follows Chapter 5 in \cite{hirsa2016computational} mostly. We split the integral term in Equation \eqref{eq:pide_middle} into two parts, the integrals on $\vert y\vert \leq \epsilon$ and $\vert y\vert > \epsilon$ respectively.

In the region $\vert y\vert \leq \epsilon$, 
$$w(x+y,\tau)=w(x,\tau)+y\frac{\partial w}{\partial x}(x,\tau)+\frac{y^2}{2} \frac{\partial^{2} w}{\partial x^{2}}(x,\tau)+O(y^{3})$$
and $$e^{y}=1+y+\frac{y^2}{2}+O(y^{3}).$$
Using those two approximations, we get  

\begin{eqnarray*}
	&& \int_{\vert y\vert \leq \epsilon}\left[w(x+y,\tau)-w(x,\tau)-\frac{\partial w}{\partial x}(x,\tau)(e^{y}-1)  \right]k(y)dy\\
	&=&\int_{\vert y\vert \leq \epsilon}\left[\frac{y^2}{2}\frac{\partial^{2} w}{\partial x^{2}}(x,\tau)-\frac{y^2}{2}\frac{\partial w}{\partial x}(x,\tau)+O(y^{3}) \right]k(y)dy\\
	&\approx & \int_{\vert y\vert \leq \epsilon}\left[\frac{y^2}{2}\frac{\partial^{2} w}{\partial x^{2}}(x,\tau)-\frac{y^2}{2}\frac{\partial w}{\partial x}(x,\tau) \right]k(y)dy.\\
\end{eqnarray*}
Define $\sigma^{2}(\epsilon)=\int_{\vert y\vert \leq \epsilon}y^{2}k(y)dy$ and we get $$\int_{\vert y\vert \leq \epsilon}\left[w(x+y,\tau)-w(x,\tau)-\frac{\partial w}{\partial x}(x,\tau)(e^{y}-1)  \right]k(y)dy \approx \frac{1}{2} \sigma^{2}(\epsilon)\left( \frac{\partial^{2} w}{\partial x^{2}}(x,\tau)-\frac{\partial w}{\partial x}(x,\tau)\right) .$$

In the region $\vert y\vert > \epsilon$, 
\begin{eqnarray*}
	&& \int_{\vert y\vert > \epsilon}\left[w(x+y,\tau)-w(x,\tau)-\frac{\partial w}{\partial x}(x,\tau)(e^{y}-1)  \right]k(y)dy\\
	&=&\int_{\vert y\vert > \epsilon}\left[w(x+y,\tau)-w(x,\tau)  \right]k(y)dy+\frac{\partial w}{\partial x}(x,\tau)\omega(\epsilon ), \\
\end{eqnarray*}
where $w(\epsilon )=\int_{\vert y\vert >\epsilon}(1-e^{y})k(y)dy$.

Combining the two parts of integrals and putting them back to Equation \eqref{eq:pide_middle}, we get
\begin{eqnarray}
	\frac{1}{2}\sigma^{2}(\epsilon)\frac{\partial^{2} w}{\partial x^{2}}(x,\tau) +\int_{\vert y\vert >\epsilon }\left[w(x+y,\tau)-w(x,\tau) \right]k(y)dy&&\notag \\
	-\frac{\partial w}{\partial \tau}(x,\tau)+(r-q+\omega(\epsilon)-\frac{1}{2}\sigma^{2}(\epsilon) )\frac{\partial w}{\partial x}(x,\tau)-rw(x,\tau)&=&0.\label{eq:pide_ap}
\end{eqnarray}
The derivative terms are calculated by back-propagation of neural networks. The integral $\int_{\vert y\vert >\epsilon }\left[w(x+y,\tau)-w(x,\tau) \right]k(y)dy$ is calculated using the trapezoidal rule. In the numerical experiments, the grid points for the trapezoidal rule are chosen to be 
\begin{align*}
	y_{j} &= \begin{cases}
		0.01j,& 1\leq j < 50,\\
		0.05(j-50) + 0.5, & 50\leq j<60,\\
		0.2(j-60) + 1, & 60\leq j < 75,\\
		-y_{-j}, & -75<j\leq -1.
	\end{cases}
\end{align*}  
The grid points are denser around 0 and coarser far from 0 because $k(y)$ decreases exponentially when $\vert y\vert$ increases. By the trapezoidal rule, we have
\begin{align*}
	&\int_{\vert y\vert >\epsilon }\left[w(x+y,\tau)-w(x,\tau) \right]k(y)dy\\
	\approx &\left(w(x+y_1,\tau)-w(x,\tau) \right)k(y_1)(y_2-y_1)/2 \\
	&+\sum_{j=2}^{73}\left(w(x+y_{j},\tau)-w(x,\tau) \right)k(y_{j})(y_{j+1}-y_{j-1})/2+ \\ 
	&+\left(w(x+y_{74},\tau)-w(x,\tau) \right)k(y_{74})(y_{74}-y_{73})/2\\
	&+\left(w(x+y_{-1},\tau)-w(x,\tau) \right)k(y_{-1})(y_{-1}-y_{-2})/2 \\
	&+\sum_{j=2}^{73}\left(w(x+y_{-j},\tau)-w(x,\tau) \right)k(y_{-j})(y_{-j+1}-y_{-j-1})/2+ \\ 
	&+\left(w(x+y_{-74},\tau)-w(x,\tau) \right)k(y_{-74})(y_{-73}-y_{-74})/2,
\end{align*}
which is a linear combination of the values of $w(\cdot, \tau)$.

\end{document}